\documentclass[twocolumn,showpacs,amsmath,amssymb,superscriptaddress,prb]{revtex4}

\usepackage{bm}
\usepackage{epsfig}
\usepackage{color}

\begin{document}


\title{Longer-range lattice anisotropy strongly competing with spin-orbit interactions in pyrochlore iridates}

\author{L. Hozoi}
\affiliation{Institute for Theoretical Solid State Physics, IFW Dresden, Helmholtzstr.~20, 01069 Dresden, Germany}
\author{H. Gretarsson}
\author{J. P. Clancy}
\affiliation{Department of Physics, University of Toronto, 60 St.~George Street, Toronto, Ontario M5S 1A7, Canada}
\author{B.-G. Jeon}
\affiliation{CeNSCMR, Department of Physics and Astronomy, Seoul National University, Seoul 151-747, Korea}
\author{B. Lee}
\affiliation{CeNSCMR, Department of Physics and Astronomy, Seoul National University, Seoul 151-747, Korea}
\author{K. H. Kim}
\affiliation{CeNSCMR, Department of Physics and Astronomy, Seoul National University, Seoul 151-747, Korea}
\author{V. Yushankhai}
\affiliation{Max-Planck-Institut f\"{u}r Physik komplexer Systeme, N\"{o}thnitzer Str.~38, 01187 Dresden, Germany}
\affiliation{Joint Institute for Nuclear Research, Joliot-Curie 6, 141980 Dubna, Russia}
\author{Peter Fulde}
\affiliation{Max-Planck-Institut f\"{u}r Physik komplexer Systeme, N\"{o}thnitzer Str.~38, 01187 Dresden, Germany}
\affiliation{POSTECH, San 31 Hyoja-dong, Namgu Pohang, Gyeongbuk 790-784, Korea}
\author{D. Casa}
\affiliation{Advanced Photon Source, Argonne National Laboratory, Argonne, Illinois 60439, USA}
\author{T. Gog}
\affiliation{Advanced Photon Source, Argonne National Laboratory, Argonne, Illinois 60439, USA}
\author{Jungho Kim}
\affiliation{Advanced Photon Source, Argonne National Laboratory, Argonne, Illinois 60439, USA}
\author{A. H. Said}
\affiliation{Advanced Photon Source, Argonne National Laboratory, Argonne, Illinois 60439, USA}
\author{M. H. Upton}
\affiliation{Advanced Photon Source, Argonne National Laboratory, Argonne, Illinois 60439, USA}
\author{Young-June Kim}
\affiliation{Department of Physics, University of Toronto, 60 St.~George Street, Toronto, Ontario M5S 1A7, Canada}
\author{Jeroen van den Brink}
\affiliation{Institute for Theoretical Solid State Physics, IFW Dresden, Helmholtzstr.~20, 01069 Dresden, Germany}
\affiliation{Department of Physics, Technical University Dresden, D-01062 Dresden, Germany}

\begin{abstract}
In the search for topological phases in correlated electron systems,
materials with $5d$ transition-metal ions, in particular,
the iridium-based pyrochlores $A_2$Ir$_2$O$_7$, provide fertile grounds.
Several novel topological states have been predicted but the actual realization
of such states is believed to critically depend on the strength of local potentials
arising from distortions of the IrO$_6$ cages.
We test this hypothesis by measuring with resonant inelastic x-ray scattering the electronic
level splittings in the $A$=Y, Eu systems, which we show to agree very well with
{\it ab initio}
quantum chemistry
electronic-structure calculations for the series of materials with $A$=Sm, Eu, Lu, and Y.
We find, however, that not distortions of the IrO$_6$ octahedra are the primary source
for quenching the spin-orbit interaction, but longer-range lattice anisotropies which
inevitably break the local cubic symmetry.
\end{abstract}

\date\today

\maketitle

\section{Introduction}
It is remarkable that the electronic bands of simple, non-interacting electron systems have intrinsic topological properties which have only recently been uncovered.\cite{TIs_Kane05,TIs_Bernevig06,TIs_Konig07}
The presence of an insulating state of topological nature has been established in, for instance, a number of bismuth based materials,\cite{TIs_Hsieh08,Zhang09,Rasche13} where this state is driven by the strong spin-orbit interaction (SOI) of the rather delocalized bismuth $6p$ electrons. These materials can be classified as either strong or weak topological insulators (TI's).\cite{Moore07,TIs_Fu07,Roy09}

The observed richness of topological states already on the single-electron level prompts the intriguing question what kind of topological phases can develop in more strongly correlated, many-body electron systems. Correlation effects, in particular, intra and inter-orbital electron-electron interactions, are very substantial in $3d$ transition-metal compounds such as the copper oxides. However, they become progressively weaker when going to heavier transition-metal elements, i.e., $4d$ and $5d$ systems, as the $d$ orbitals become more and more extended. Yet the relativistic SOI, the root cause of a number of many topologically non-trivial electronic states,  follows the opposite trend -- it increases progressively when going from $3d$ to $5d$ elements. In $5d$ transition-metal compounds like the iridates, the interesting situation arises that the SOI and Coulomb interactions meet on the same energy scale. The electronic structure of iridates therefore depends on a strong competition between the electronic hopping amplitudes, local energy-level splittings, electron-electron interaction strengths, and the SOI of the Ir $5d$ electrons. It is very interesting that the interplay of these ingredients in principle allows the stabilization of entirely novel electronic states such as strong or weak topological Mott states, an axion insulator or a Weyl semi-metallic state.\cite{227_guo_2009,227Ir_pesin_2010,227Ir_savrasov_2011,227Ir_balents_2011,227Ir_witczak_2012}

\begin{figure}
\includegraphics[width=0.61\columnwidth]{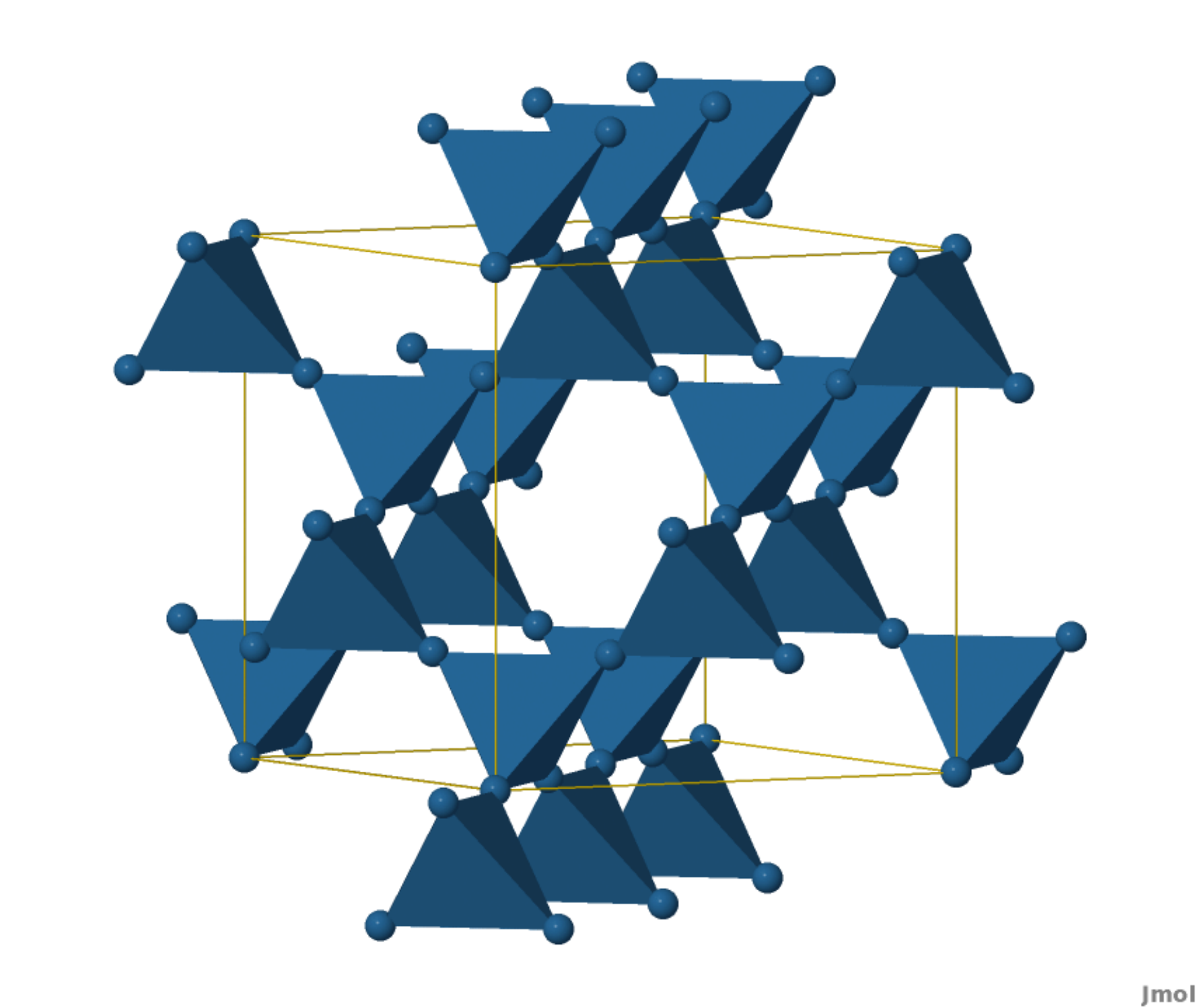}
\includegraphics[width=0.37\columnwidth]{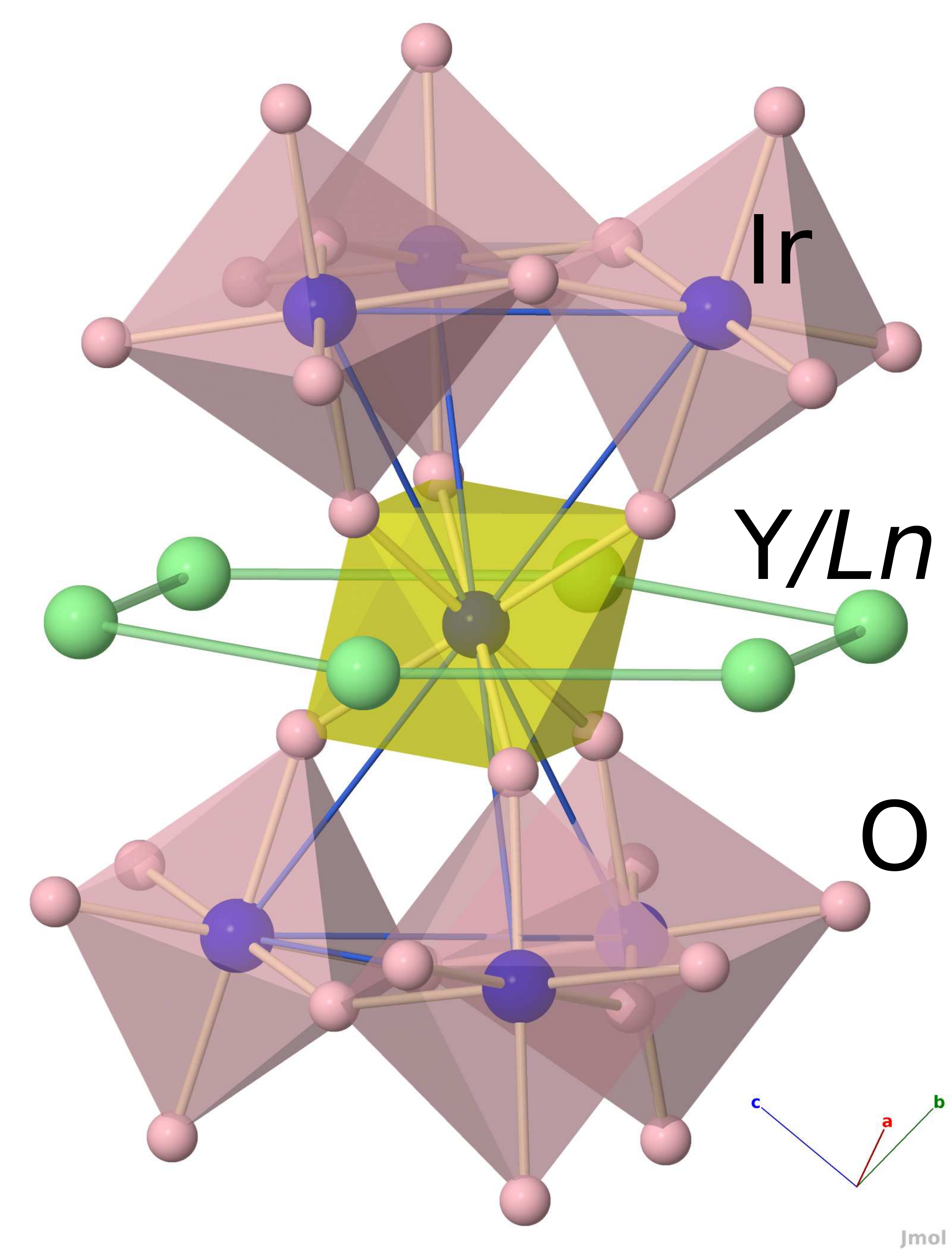}
\caption{
(a) Network of corner-linked Ir$_4$ tetrahedra in $A_2$Ir$_2$O$_7$ pyrochlore iridates.
(b) A central IrO$_6$ octahedron
and how it is connected to its six neighboring IrO$_6$ octahedra.
The six trivalent $A$-site atoms
form a hexagonal ring around the central octahedron.
}
\label{fig:pyrochlore}
\end{figure}

In the pyrochlore iridates of the type $A_2$Ir$_2$O$_7$ that we consider here, with $A$\,=\,Sm, Eu, Lu, and Y, five electrons occupy the three Ir $t_{2g}$ orbitals, which reside at Ir$^{4+}$ sites inside corner-linked IrO$_6$ octahedra, see Fig.~\ref{fig:pyrochlore}. This leaves one hole in the $t_{2g}$ shell to which thus 6 distinct $t_{2g}$ quantum states (3 orbital and 2 spin) are available. When the local symmetry is cubic, so that it does not lift the degeneracy of the three $t_{2g}$ levels, the strong SOI splits the $t_{2g}$ states up into a pure $j\!=\!3/2$ quadruplet and a pure $j\!=\!1/2$ doublet. The doublet is higher in energy and therefore accommodates the hole. Any additional crystal-field splitting, for instance of tetragonal or trigonal symmetry, lifts the degeneracy of $t_{2g}$ states and competes with the spin-orbit coupling, thus tending to quench the orbital moment. As the SOI is driving the formation of electronic states of topological nature, the outcome of this competition is decisive for the actual realization of any type of nontrivial topological ground state in pyrochlore iridates.\cite{227Ir_baek_kim_10,227Ir_kargarian_11}

\section{Experimental RIXS results}
We use Resonant Inelastic X-ray Scattering (RIXS)~\cite{Ament11a} to
measure directly the energy of the different configurations of a hole in the Ir $t_{2g}$
shell~\cite{Ament11b}
of Y$_2$Ir$_2$O$_7$ and Eu$_2$Ir$_2$O$_7$ and to determine that way the crystal-field energy splittings of these states.
The single crystals of Eu$_2$Ir$_2$O$_7$ and powder samples of Y$_2$Ir$_2$O$_7$ were grown by solid-state
synthesis.
Mixtures of Y$_2$O$_3$ and IrO$_2$ with purities of 99.99\%
were ground in stoichiometric molar ratios, pelletized,
and then heated in air at 1000$^\circ$C for 100 hours.
The resulting material was reground, pressed into pellets, and resintered at the same
temperature for an additional 150 hours, with two intermediate regrindings.
Powder X-ray diffraction measurements confirmed the phase purity of
the resulting Y$_2$Ir$_2$O$_7$ sample to within the resolution of the measurement.
A single crystal of Eu$_2$Ir$_2$O$_7$ was grown by the solid-state synthesis method, as previously described in detail in
Ref.~\onlinecite{crystals_227_millican}.
A mixture of polycrystalline Eu$_2$Ir$_2$O$_7$ and KF (2N) were heated up to 1100$^\circ$C and next cooled down to 850$^\circ$C at
a rate of 2.5$^\circ$/h.
Resistivity data on the resulting Eu$_2$Ir$_2$O$_7$ single crystal shows almost metallic behavior,
indicating that the sample displays slight excess of Ir, see the discussion in Ref.~\onlinecite{crystals_227_ishikawa}.

RIXS is a second-order scattering technique and can directly probe the electronic transitions
within the Ir $5d$ manifold due to two successive electric dipole transitions ($2p\!\rightarrow\!5d$
followed by $5d\!\rightarrow\!2p$).\cite{Ament11a,Ament11b}
It is therefore a valuable technique for detecting transitions between crystal-field split Ir $5d$ levels
and has been utilized for a variety of iridates.\cite{Kim12,CuIr2S4_gretarsson11,214Ir_ishii11,Liu12,213_rixs_gretarsson_2012}
We determine the splittings by measuring the $d$-$d$ transition energies at the iridium $L_3$ edge, with
an incident energy, $E_i\!=\!11.217$ keV, chosen to maximize the resonant enhancement of the spectral features of interest below 1.5~eV.
It should be noted that varying the incident energy did not result in a shift of any of the
peaks but merely changed their intensities, a behavior associated with valence excitations.
\cite{Kim12,CuIr2S4_gretarsson11,214Ir_ishii11,Liu12,213_rixs_gretarsson_2012}
The experiments were carried out at the Advanced Photon Source using the 9ID beamline with a 
\textcolor{black}
{Si(444)}
channel-cut secondary monochromator and a horizontal scattering geometry. A spherical (1~m radius) diced
Si(844) analyzer was used and an overall energy resolution of
\textcolor{black}
{
175
}
meV (FWHM) was obtained. 
\textcolor{black}
{
Higher resolution measurements were carried out using the MERIX spectrometer on beamline 30-ID-B.  Measurements were performed using a spherical (2~m radius) diced Si(844) analyzer and a channel-cut Si(844) secondary monochromator to give an overall energy resolution (FWHM) of 35 meV.
}

Due to experimental conditions, the spectra for Y$_2$Ir$_2$O$_7$ and Eu$_2$Ir$_2$O$_7$ were
obtained at two different temperatures, 300 and 150 K, respectively.
Since the thermal contraction of Y$_2$Ir$_2$O$_7$ is extremely small, within tenths of a
percentage between 300 and 150 K, and the local oxygen octahedra are unaffected by the temperature,
\cite{Ir227_mg_order_11} we can conclude that this difference in temperature has minimum bearing
on our results.
The RIXS spectra of both Y$_2$Ir$_2$O$_7$ and Eu$_2$Ir$_2$O$_7$ in Fig.~2 show sharp
features below 1.5 eV, corresponding to transitions within the Ir $t_{2g}$ levels,
and a strong intense peak stretching from 2 to 5 eV that according to the
calculations, see below, corresponds to $d$-$d$ transitions between the Ir $t_{2g}$
and $e_g$ levels.
To quantitatively analyze the RIXS spectra, the various peaks were fitted with analytical functions,
as shown by the dashed lines in Fig.~2.
The low-energy excitations, $E_1$ and $E_2$, was fitted with one Gaussian and one Lorentzian,
respectively.
The Lorentzian function was used in order to capture some of the tail on the high energy side ($\sim\!1.5$ eV).
The high-energy  excitations ($E_3$) were fitted with a Gaussian peak on top of a sloping background.
Such a sloping background may come from charge transfer excitations which are expected to
appear in this range.

The position refinement for the three main peaks apart from the zero-loss
peak results in $E_1\!=\!0.53 \pm 0.05$ ($0.59 \pm 0.03$), $E_2\!=\!0.98 \pm 0.05$
\textcolor{black}
{
($0.97 \pm 0.03$), 
}
and $E_3\!=\!3.90 \pm 0.05$ eV ($3.70 \pm 0.05$ eV) for Y$_2$Ir$_2$O$_7$
(Eu$_2$Ir$_2$O$_7$).
Note that the peak widths are significantly broader than the instrumental resolution, which is about 175~meV. To check this, we carried out an additional measurement on the same Eu$_2$Ir$_2$O$_7$ sample employing much higher energy resolution of 35~meV. This high resolution spectrum is overlayed on top of the low resolution data in Fig.~2. As expected, the improved resolution reveals a little bit sharper features, but still with intrinsic spectral width of about 300~meV. 
\textcolor{black}
{
We note that the fitted $E_2$ peak position of the high resolution data, $0.95 \pm 0.01$~eV, is slightly smaller than that of the low resolution data, but still within the experimental error bar. Since the high resolution result has smaller error bar, this value is quoted in Table~I.
}
We also observed no noticeable momentum dependence for Eu$_2$Ir$_2$O$_7$, justifying the use of Y$_2$Ir$_2$O$_7$ powder samples for our comparison.

\begin{figure}
\includegraphics[width=0.9\columnwidth]{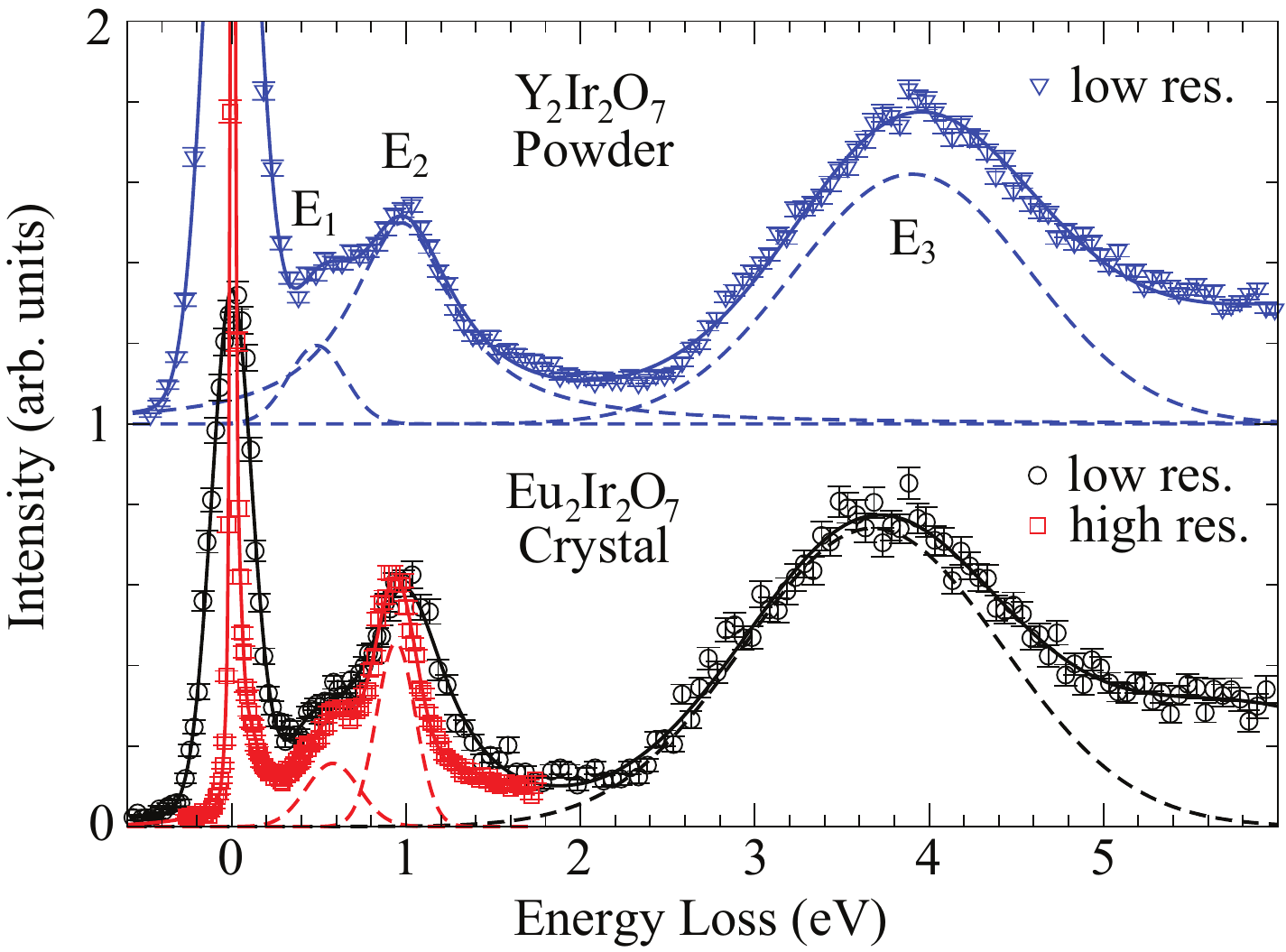}
\caption{Resonant inelastic X-ray scattering spectra of Y$_2$Ir$_2$O$_7$ (top) and Eu$_2$Ir$_2$O$_7$ (bottom) at the iridium $L$-edge. The Eu$_2$Ir$_2$O$_7$ spectra were collected at $\mathbf{Q}$=(8.45,8.45,7). For Eu$_2$Ir$_2$O$_7$, high resolution data obtained for low energy region at (7.8,7.8,7.8) are also plotted to emphasize the intrinsic nature of the peak width. Dashed curves are the result of fitting (see text).
}
\label{RIXS}
\end{figure}

\section{Empirical description of RIXS results}

Using a first, empirical ansatz we fit the energies of the low-lying $E_1$ and $E_2$ peaks to the eigenvalues of an
effective single-ion Hamiltonian for the $t_{2g}$ orbitals of the form
$H_0\!=\!\lambda  \bm{l} \cdot \bm{s} -\Delta  l_z^2$, where $\lambda $ is the SOI strength and
$\Delta $ the $t_{2g}$ crystal-field splitting.\cite{book_abragam_bleaney,SOC_d5_thornley68,Liu12}
The latter tends to quench the Ir orbital moment and is usually identified with distortions of the IrO$_6$ octahedra.
\cite{213Ir_jackeli_09,227Ir_baek_kim_10}
The eigenvalues of $H_0$ and the splittings among the spin-orbit $t_{2g}^5$ states are given
by the following expressions:
$E_0\!=\!\lambda(-1+\delta-\sqrt{9+2\delta+\delta^2})/4$,
$E_1\!=\!\lambda/2$,
and
$E_2\!=\!\lambda(-1+\delta+\sqrt{9+2\delta+\delta^2})/4$,
where $E_0$ is the energy of the ground-state spin-orbit $j\!=\!1/2$ doublet,
$E_1$ and $E_2$ define the eigenvalues of the split $j\!=\!3/2$\,-like terms, and
$\delta\!=\!2\Delta/\lambda$.
If $E_1\!-\!E_0$ and $E_2\!-\!E_0$ are known from experiment, i.e., RIXS, simple estimates
for $\lambda$ and $\Delta$ can be in principle derived from the relations above.
In particular,
$\lambda\!=\!2(2E_{10}\!-\!E_{20})/(3\!-\!\delta)$ and
$\Delta\!=\!\lambda\delta /2$, where
$\delta\!=\!-b\!-\!\sqrt{b^2\!-\!9}$,
$b\!=\!(1\!+\!3a^2)/(1\!-\!a^2)$,
$a\!=\!E_{20}/(E_{20}\!-\!2E_{10})$,
$E_{20}\!=\!E_2\!-\!E_0$, and
$E_{10}\!=\!E_1\!-\!E_0$\,.
The fit of the RIXS data to such a $\lambda$--$\Delta$ model yields the effective parameters
$\lambda \!=\!0.43$ and $\Delta \!=\!0.56$ eV for Y-227 and
$\lambda \!=\!0.46$ and $\Delta \!=\!0.46$ eV for Eu-227.
The value of $\lambda $ for each of these materials agrees with values of 0.39--0.49 eV
extracted from electron spin resonance measurements on Ir$^{4+}$ impurities.
\cite{lambda_Ir_andlauer76}
The magnitude of $\Delta $, 0.46--0.56 eV, however, is surprisingly large.
To understand the size and elucidate the microscopic origin of this large crystal-field
splitting --\,a crucial energy scale in determining the topological ground state of the
electronic system\,-- we have carried out a set of detailed {\it ab initio} calculations of the
Ir $d$-level electronic structure on a series of $5d^5$ pyrochlore iridates: Sm-, Eu-, Lu-, and Y-227.

\begin{table}
\caption{
Relative energies (in eV) of the split $j\!=\!3/2$ states $E_1$ and $E_2$
as well as $t_{2g}^5$ to $t_{2g}^4e_g^1$ excitation energies in 227 iridates.
The results are the outcome of {\it ab initio} spin-orbit MRCI calculations,
see text.
The experimental values for Eu$_2$Ir$_2$O$_7$ and Y$_2$Ir$_2$O$_7$ determined from their RIXS spectra in Fig.\,2 are shown in bold, within brackets.
}
\begin{ruledtabular}
\begin{tabular}{llll}
                    &$E_1$        &$E_2$        &$t_{2g}^4e_g^1$   \\
\hline
\\
Sm$_2$Ir$_2$O$_7$   &0.61         &0.91         &3.41--4.75        \\
Eu$_2$Ir$_2$O$_7$   &0.60 ({\bf 0.59})  &0.91 ({\bf 0.95})  &3.39--4.72 ({\bf 3.70}) \\
Lu$_2$Ir$_2$O$_7$   &0.57         &0.92         &3.49--4.88        \\
 Y$_2$Ir$_2$O$_7$   &0.58 ({\bf 0.53})  &0.94 ({\bf 0.98})  &3.48--4.84 ({\bf  3.90}) \\
\end{tabular}
\end{ruledtabular}
\label{dd_SOC}
\end{table}

\section{{\it Ab initio} calculation of d-d excitations}

To investigate in detail the electronic structure and the essential interactions in the $A_2$Ir$_2$O$_7$
iridates, we rely on {\it ab initio} many-body techniques from wave-function-based quantum chemistry.
\cite{bookQC_2000}
Multiconfiguration self-consistent-field (MCSCF) and multireference configuration-interaction
(MRCI) calculations \cite{bookQC_2000} were carried out to this end on properly embedded
finite clusters.
Since it is important to accurately describe the charge distribution at sites in the immediate
neighborhood,\cite{qc_NNs_degraaf_99,CuO2_dd_hozoi11,CuO_ZRB_hozoi_07} we explicitly include in the
actual cluster the closest six $A$-ion neighbors and the six adjacent IrO$_6$ octahedra
around the reference IrO$_6$ unit for which the Ir $d$-$d$ excitations are explicitly
computed, see also
Refs.~\onlinecite{Liu12,213_rixs_gretarsson_2012,214Ir_vmk_2012,113Ir_bogdanov_2012,Os227_bogdanov_12}.
The solid-state surroundings were further modeled as a large array of point charges fitted
to reproduce the crystal Madelung field in the cluster region.
All calculations were performed with the {\sc molpro} quantum chemistry software.
\cite{molpro_brief}

We used energy-consistent relativistic pseudopotentials for Ir and the $A$ elements
\cite{ECP_5d_stoll,ECP_4f_dolg89,ECP_4f_dolg93,ECP_4d_peterson07} and Gaussian-type valence
basis functions.
Basis sets of quadruple-zeta quality were applied for the valence shells of the central
Ir$^{4+}$ ions \cite{ECP_5d_stoll} and triple-zeta basis sets
for the ligands \cite{BS_O_VTZ} of the central octahedron and for the nearest-neighbor (NN)
Ir sites.\cite{ECP_5d_stoll}
For the central Ir ions we also used two polarization $f$ functions.\cite{ECP_5d_stoll}
For farther O's around the NN Ir sites we applied minimal atomic-natural-orbital basis sets.
\cite{ANOs_pierloot_95}
The $f$ electrons of the $Ln^{3+}$ species were incorporated in the effective core potentials
\cite{ECP_4f_dolg89,ECP_4f_dolg93} and the outer $sp$ shells of the $Ln^{3+}$ and Y$^{3+}$ ions were
modeled with sets of $[3s2p]$ functions.\cite{ECP_4f_dolg89,ECP_4f_dolg93,ECP_4d_peterson07}
Crystallographic data as reported by Taira {\it et al.}\cite{227Ir_taira_2001} were employed.

For the ground-state calculations, the orbitals within each finite cluster are variationally optimized
at the MCSCF level.
All Ir $t_{2g}$ functions are included in the active orbital space \cite{bookQC_2000}, i.e.,
all possible electron occupations are allowed within the $t_{2g}$ set of orbitals.
On-site $t_{2g}$ and $t_{2g}$ to $e_g$ excitations are afterwards computed just for the
central IrO$_6$ octahedron while the occupation of the NN Ir valence shells is held frozen
as in the ground-state configuration.
The MRCI treatment includes on top of the MCSCF wave functions single and double excitations
\cite{bookQC_2000} from the O $2p$ orbitals at the central octahedron and the Ir $5d$ orbitals.

To extract the local Ir $t_{2g}$ splittings, the NN Ir$^{4+}$ $d^5$ ions were explicitly
included in a first set of MCSCF calculations.
However, the presence of six open-shell Ir NN's makes the spin-orbit calculations cumbersome
because for seven $5d^5$ sites (and nondegenerate orbitals), a given electron configuration
implies 1 octet, 6 sextet, 14 quartet, and 14 doublet states which further interact via
spin-orbit coupling.
To simplify the problem and reduce the computational effort, we therefore further replaced the
six Ir$^{4+}$ $d^5$ NN's by closed-shell Pt$^{4+}$ $d^6$ ions
\footnote{
This is an usual procedure in quantum chemistry studies on transition-metal systems, see, e.g.,
Refs.~\onlinecite{Os227_bogdanov_12,qc_NNs_degraaf_99,Na2V2O5_hozoi_02,SIA_Fe_maurice_2013}.
}
and in this manner obtained
the relative MRCI energies for the spin-orbit states presented in Tables I and II.

The $d$-$d$ splittings calculated for cuprates such as
La$_2$CuO$_4$ and Sr$_2$CuO$_3$ \cite{CuO2_dd_hozoi11,spin_orb_sep_schlappa12,CuO2_dd_huang11}
and iridates such as Sr$_2$IrO$_4$, Na$_2$IrO$_3$, and Sr$_3$CuIrO$_6$ \cite{214Ir_vmk_2012,Liu12,213_rixs_gretarsson_2012}
by similar quantum chemistry techniques are in close agreement to the
experimental values of these excitation energies.
Also for the 227 iridium pyrochlores that we consider here, we observe that the calculated
excitation energies of the 5$d$ multiplets (the values of $E_1$, $E_2$, and $E_3$), see Table I,
are in close agreement with the ones obtained from our RIXS experiments.

\section{Discussion}

The good agreement between calculated and measured $d$-$d$ excitation energies forms the
basis for a subsequent detailed analysis of the microscopic origin of the crystal-field splitting
of the Ir 5$d$ levels.
To this end, we first test the hypothesis that the splitting $\Delta$ in the effective single
ion $\lambda\!-\!\Delta$ model is due to a distortion of the IrO$_6$ octahedra which lowers the
local cubic symmetry to trigonal (or even lower) symmetry.
It turns out that the crystal structure of the 227's under consideration is fully defined by
just three parameters: the space-group number, the cubic lattice constant $a$, and the fractional
coordinate $x$ of the O at the $48f$ site.\cite{227Ir_taira_2001}
For $x\!=\!x_c\!=\!5/16$, the oxygen cage around each Ir site forms an undistorted, regular
octahedron.
In our 227 Ir pyrochlores, however, $x$ is always larger than $x_c$, which translates into a
compressive trigonal distortion of the IrO$_6$ octahedra and hence a splitting of the $5d$ electronic levels.
To estimate how large the resulting trigonal crystal-field splitting is, we have performed a
set of further {\it ab initio} calculations, but now for an idealized crystal structure
with $x\!=\!x_c$ and thus undistorted octahedra.
The results of the spin-orbit calculations listed in Table II show that the $j\!=\!3/2$\,-like
states are split off by a sizable amount even for $x\!=\!x_c$.
Thus local trigonal distortions of the IrO$_6$ octahedra are {\it not} the main cause of the energy splitting
$\Delta$ of the Ir $t_{2g}$ levels.

\begin{table}
\caption{
Calculated energies, $E^0_1$,  $E^0_2$, of the $j\!=\!3/2$\,-like spin-orbit states
in idealized crystal structures without trigonal distortion of the IrO$_6$ octahedra.
Those states are split even without trigonal squashing.
$\bar{\Delta}^0$ and $\bar{\Delta}$ are the splitting of the Ir $t_{2g}$ levels in
MRCI calculations without SOI in the undistorted idealized and experimental\cite{227Ir_taira_2001} crystal structures, respectively.
}
\begin{ruledtabular}
\begin{tabular}{ccc|cc}
                    &\multicolumn{2}{c}{Undistorted Octahedron}  & \multicolumn{2}{c}{Without SOI} \\
\hline
                    &$E^0_1$  &$E^0_2$   &$\bar{\Delta}^0$ & $\bar{\Delta}$ \\
\hline
Eu$_2$Ir$_2$O$_7$   &0.67   &0.89    &0.30  &0.27        \\
 Y$_2$Ir$_2$O$_7$   &0.66   &0.90    &0.32  &0.30        \\
\end{tabular}
\end{ruledtabular}
\label{dd_xc}
\end{table}

To understand the physical origin of the large Ir $t_{2g}$ splittings one needs to consider
the crystal structure of the $A$-227's in more detail.
As shown in Fig.~1b, the $A$ ions closest to a given Ir site form a hexagonal structure
in a plane parallel to two of the facets of the IrO$_6$ octahedron.
Even without pushing those two facets of the octahedral cage closer to each other, the six
adjacent $A$ cations generate a trigonal field that breaks cubic symmetry.
In the simplest picture this positive potential stabilizes the Ir $e_{g}^{\prime}$ orbitals.
The latter have the electronic charge closer to the plane defined by the six $A$ NN's,
as compared to the $a_{1g}$ orbital.
There is however a competing effect related to the positive NN Ir ions, three above and three
below the plane of adjacent $A$ sites, see Fig.~1.
The potential generated by the Ir NN's stabilizes the $a_{1g}$ sublevel.
The numerical results we obtain, see below, indicate that
the net effect of this anisotropic arrangement of the nearby $A$ and Ir cations on the
$t_{2g}$ level splitting is stronger than the effect of the trigonal distortion of the
IrO$_6$ cages.
It is interesting to note that similar effects are expected in pyrochlore systems with $3d$
transition-metal ions.   
However, the $3d$ orbitals being more localized will reduce the effect of such anisotropies
beyond the first ligand coordination shell and result in smaller splittings of the energy
levels.

MRCI calculations without SOI's, see Table II, show that the magnitude of the $t_{2g}$ splittings
is about the same in the distorted, experimental crystal structure ($\bar{\Delta}$, $x\!>\!x_c$) and the
idealized, undistorted structural model ($\bar{\Delta}^0$, $x\!=\!x_c$).
This confirms that the splitting of the Ir $t_{2g}$ levels is due to anisotropic potentials beyond
the NN ligand coordination shell.
The role of nearby cations in generating anisotropic fields that compete with the trigonal
distortion of the ligand cage has been pointed out as early as the 60's for the spinel structure,
\cite{trig_fields_slonczewski_61,trig_fields_berger_65} recently confirmed by {\it ab initio}
quantum chemistry calculations on the $S\!=\!3/2$ 227 pyrochlore Cd$_2$Os$_2$O$_7$,\cite{Os227_bogdanov_12}
and also analyzed for layered Co oxide compounds by density-functional calculations.\cite{NaCoO2_pillay_08}

The calculations on the 227 iridates also show that the $a_{1g}$ sublevel is lower in energy than
the $e_{g}^{\prime}$ sublevels, which is usually referred to as negative trigonal splitting.
\cite{d3_tanabe_sugano_58,NaCoO2_pillay_08}
This indicates that the effect of the positive potential related to the six Ir NN's is stronger
than the effect of the field generated by the closest $A$ ions.
A stabilization of the $a_{1g}$ orbital due to positive ions on the trigonal axis has been
earlier evidenced by Pillay {\it et al.} in Na$_x$CoO$_2$.\cite{NaCoO2_pillay_08}
Importantly, if in the quantum chemistry calculations the nuclear charge is artificially lowered by 1 at each
of the six Ir NN sites and raised by 1 at each of the six $A$ NN sites, the trigonal splitting
changes sign.

While the {\it ab initio} calculations without SOI yield $\Delta$ values of $\approx$0.30 eV,
the fit of the RIXS data with the effective $\lambda\!-\!\Delta$ model provides $t_{2g}$ splittings
$\Delta$ of 0.46--0.56 eV, more than 50$\%$ larger.
This indicates that in pyrochlore iridates the effect of the relativistic spin-orbit coupling
cannot be completely captured by oversimplified models such as the $\lambda\!-\!\Delta$
Hamiltonian.
Additional degrees of freedom must be considered for the construction of a minimal effective
model, i.e., hybridization effects, Ir--O and Ir $e_{g}^{\prime}$--$e_{g}$, in the
presence of trigonal external fields and also many-body $d$-shell correlations.
The $e_{g}^{\prime}$--$e_{g}$ couplings, for instance, were found to be important in
trigonally distorted $3d^5$ compounds.\cite{NaCoO_lepetit_08}
Our data in Table II, showing that with trigonal distortions the $a_{1g}$--$e_{g}^{\prime}$
splittings decrease
\textcolor{black}
{
(as $\bar{\Delta} < \bar{\Delta}^0$),
}
qualitatively confirm the quantum chemistry results of Landron and Lepetit,\cite{NaCoO_lepetit_08} i.e.,
in the presence of trigonal squashing the $e_{g}^{\prime}$ levels are energetically favored as compared
to the $a_{1g}$ states, in contrast to naive expectations based on one-electron crystal-field theory.
Obviously, for $5d$ oxides, the $t_{2g}$--$e_g$ couplings are further enhanced by the strong SOI's.
\cite{SOC_d5_thornley68,214Ir_haskel_2012}

It is also interesting that without trigonal distortions the lower doublet state
originating from the $j\!=\!3/2$ quartet significantly shifts to higher energy as compared to the
trigonally compressed experimental structure
\textcolor{black}
{
($E_1 < E_1^0 $, see Tables II and I),
}
although the splitting of the $t_{2g}$ levels is about the same in the two cases.
The other doublet at somewhat higher energy is on the other hand not much affected.
A relevant detail is here that the Ir-O bond lengths are slightly reduced for the data in
Table II because the lattice constant was kept the same and in order to remove the trigonal
distortion only the fractional coordinate $x$ of the O site was modified.
Shorter Ir-O bonds yield higher electron density at the Ir site.
The $5d$-shell Mulliken population,\cite{bookQC_2000} for instance, is larger by 0.1 of an
electronic charge for the structural model without trigonal distortions.
The results of the spin-orbit calculations in Table I and Table II, with a sizable shift of
the $E^0_1$ level to higher energy, indicate that the charge redistribution within
the IrO$_6$ octahedron and the higher electron density at the Ir site effectively modify
the spin-orbit couplings within the IrO$_6$ unit.
Feeding the $E^0_1$ and $E^0_2$ quantum chemistry results of Table II to a simple $\lambda$--$\Delta$
model yields indeed a rather large $\lambda$ effective parameter of 0.49--0.50 while the corresponding
$\Delta$'s perfectly match this time the {\it ab initio} trigonal splittings $\bar{\Delta}^0$ computed with
no trigonal distortions, 0.30 and 0.32 eV.
\textcolor{black}
{
Thus in an idealized system, without trigonal distortions, the effective $\lambda$--$\Delta$ model provides a reasonable description.
}
It is, however, clear that additional ingredients are required in the effective model for a qualitative
description of, e.g., how those excitation energies evolve with the amount of trigonal distortion.
This will be the topic of future investigations.

\section{Conclusions}
We have presented here both experimental and theoretical evidence for the presence of large Ir $t_{2g}$ splittings in pyrochlore iridates. These splittings arise from longer-range crystal anisotropies that directly compete with spin-orbit interactions. The canonical view that only local distortions of IrO$_6$ octahedra tend to quench the spin-orbit coupling in iridium compounds, in particular iridium pyrochlores, is therefore incomplete.  The broader ramification is that the rather extended nature of the $5d$ wave functions renders the longer-range anisotropy fields to be of fundamental importance throughout the $5d$ transition-metal series. Their physical effect is particularly striking when the local ligand-field symmetry is high, e.g., cubic or close to cubic, but the point-group symmetry in the crystal is lower. This includes, for example, the osmium-based pyrochlore materials.\cite{Os227_shinaoka_2012,Os227_bogdanov_12}
Yet it is pertinent for many more crystal structures, in particular, for layered quasi-2D perovskites or chain-like quasi-1D $5d$ transition-metal systems.\cite{113Ir_bogdanov_2012,Liu12}

\section{Acknowledgements}
We thank N.~A.~Bogdanov, V.~M.~Katukuri, and H.~Stoll for fruitful discussions.
Research at the University of Toronto was supported by the NSERC, CFI, and OMRI.
Use of the Advanced Photon Source at Argonne National Laboratory was supported by the U.S.~Department of Energy under Contract No.~DE-AC02-06CH11357.
Work at SNU was supported by the National CRI (2010-0018300) program.
L.~H. acknowledges financial support from the German Research Foundation (Deutsche Forschungsgemeinschaft, DFG).

\end{document}